\documentclass[fleqn,usenatbib]{mnras}

\usepackage{mathptmx}
\usepackage[T1]{fontenc}

\DeclareRobustCommand{\VAN}[3]{#2}
\let\VANthebibliography\thebibliography
\def\thebibliography{\DeclareRobustCommand{\VAN}[3]{##3}\VANthebibliography}

\usepackage{graphicx}	% Including figure files
\usepackage{ae,aecompl}
\usepackage{mathtools}
\usepackage{bm}
\usepackage{cleveref}
\usepackage{verbatim} %In the preamble
\usepackage{caption}
\usepackage{subcaption}
\usepackage{amsmath,amssymb}
\usepackage{lscape}
\usepackage{natbib}
\usepackage{rotating}
\usepackage{multirow}
\usepackage{hyperref}

\title[MOCCA in AMUSE framework: embedded gas study]{A Monte Carlo study of early gas expulsion and evolution of star clusters: new simulations with the MOCCA code in the AMUSE framework.}

\author[A. Leveque et al.]{
A.~Leveque$^{1}$\thanks{E-mail:agostino@camk.edu.pl}, 
M.~Giersz$^{1}$,
S.~Banerjee$^{2,3}$,
E.~ Vesperini$^{4}$,
J.~Hong$^{5,6}$,
\newauthor
S.~Portegies  Zwart$^{7}$
\\
$^{1}$ Nicolaus Copernicus Astronomical Center, Polish Academy of Sciences, ul. Bartycka 18, PL-00-716 Warsaw, Poland\\
$^{2}$ Helmholtz-Instituts f\"{u}r Strahlen- und Kernphysik (HISKP), Nussallee 14-16, D-53115 Bonn, Germany\\
$^{3}$ Argelander-Institut f\"{u}r Astronomie (AIfA), Auf dem\''{u}gel 71, D-53121, Bonn, Germany\\
$^{4}$ Department of Astronomy, Indiana University, Bloomington, Swain West, 727 E. 3rd Street, IN, 47405, USA\\
$^{5}$ Department of Astronomy, Yonsei University 50 Yonsei-Ro, Seodaemun-Gu, Seoul 03722, Republic of Korea\\
$^{6}$ Korea Astronomy and Space Science Institute, Daejeon 34055, Republic of Korea\\
$^{7}$ Leiden Observatory, Leiden University, PO Box 9513, 2300 RA, Leiden, The Netherlands
}
\date{Accepted XXX. Received YYY; in original form ZZZ}

\pubyear{2022}

\begin{document}
\label{firstpage}
\pagerange{\pageref{firstpage}--\pageref{lastpage}}
\maketitle

\begin{abstract}
%The abstract should briefly describe the aims, methods, and main results of the paper.
%It should be a single paragraph not more than 250 words (200 words for Letters).
%No references should appear in the abstract.
We introduce a new prescription for the evolution of globular clusters (GCs) during the initial embedded gas phase into a Monte Carlo method. 
With a simplified version of the Monte Carlo MOCCA code embedded in the AMUSE framework, we study the survival of GCs after the removal of primordial gas. 
We first test our code and show that our results for the evolution of mass and Lagrangian radii are in good agreement with those obtained with N-body simulations.
 The Monte Carlo code enables a more rapid exploration of the evolution of systems with a larger number of stars than N-body simulations.
 We have carried out a new survey of simulations to explore the evolution of globular clusters with up to $N = 500000$ stars for a range of different star formation efficiencies and half-mass radii.
 Our study shows the range of initial conditions leading to the clusters' dissolution and those for which the clusters can survive this early evolutionary phase.

\end{abstract}

\begin{keywords}
globular clusters: general - methods: numerical
\end{keywords}

%%%%%%%%%%%%%%%%%%%%%%%%%%%%%%%%%%%%%%%%%%%%%%%%%%

%%%%%%%%%%%%%%%%% BODY OF PAPER %%%%%%%%%%%%%%%%%%

\section{Introduction}
Globular clusters (GC) form through the gravitational collapse of giant molecular clouds \citep{Lada2003,Longmore2014}. Newborn clusters are then supposed to be embedded in the leftover gas. The star formation efficiency (SFE), defined as $\epsilon = M_{cl}/(M_{cl} + M_{gas})$, with $M_{cl}$ as the GC star mass and $M_{gas}$ being the embedded gas mass, represents the fraction of gas that is converted into stars.

The ultraviolet (UV) radiation of massive stars and their stellar winds and supernova explosions can lead to the expulsion of primordial gas. Indeed, the UV radiation ionizes the gas, leading to efficient coupling of the stellar radiation \citep{Hills1980,Krumholz2009}, and then to the unbinding and removal of the gas from the cluster. The radiative gas expulsion can be faster than the crossing time of the embedded gas, taking place at the sound speed for ionized hydrogen, that is $\sim 10 \,\,km/s$  \citep{Kroupa2001b,Banerjee2013}. The cluster would then expand over its dynamical timescale.  This can be crucial for the survival of the system, which may possibly be dissolved.  Different studies have been conducted in order to understand and estimate the importance of the gas expulsion, together with the response of the embedded star cluster \citep{Lada1984,Adams2000,Geyer2001,Bastian2006, Baumgardt2007,Pelupessy2012, Banerjee2013, Banerjee2014, Banerjee_2018b,Lewis2021}. Due to the gas removal phase and its consequential mass loss, the surviving cluster will have a final half-mass radius, $R_h$, 3 or 4 times  larger than its initial value \citep{Lada1984, Baumgardt2007}. In general, the expansion process and final half-mass radius value depend on several factors, e.g., SFE, gas and star density profiles, gas expulsion timescale.

Due to the small time scale of the embedded gas phase ($\sim 1-2\,\,Myr$) and the small time scale of gas expulsion ($\sim 0.1\,\,Myr$), previous studies have been carried out with NBODY codes only, because they properly account for fast global changes in the potential during the gas removal and violent relaxation phases. In this paper, we introduced and studied the embedded gas removal phase with a Monte Carlo based code. Due to the computational requirements, N-body simulations are limited to small N, with dense star clusters of millions of stars being a computational challenge \citep{Makino2003,Gaburov2009,Heggie2014,Wang2016}. On the other hand, large N and dense clusters can be simulated using a Monte Carlo code. Indeed, a great advantage of the Monte Carlo method is it provides detailed and fast dynamical evolution of GCs \citep[and references therein]{Joshi2000,Rodriguez2021,Giersz1998,Giersz2013,Giersz2019}.

With a simplified version of the MOCCA code \citep{Hypki2013}, in the AMUSE \citep{Pelupessy2013,PortegiesZwart2009,PortegiesZwart2013,PortegiesZwart2018} framework, we performed a preliminary study on the importance of the embedded gas phase to the survival of the cluster even for large number of particles.

Our paper is organized as follows. In \hyperref[sec:Methods]{section 2} and in \hyperref[sec:gasEvo]{section 3}, we introduce the methodology used in this study. In \hyperref[sec:SimAndRes]{section 4}, we present the main results, and in \hyperref[sec:Conclusions]{section 5} we finally present our discussion and conclusions. In \hyperref[Appendix]{Appendix A}, we describe the new updated version of the McLuster code used in this paper to generate the initial conditions for the studied models.

\section{Methods} \label{sec:Methods}
In this paper, we present a simplified version of the MOCCA code \citep{Giersz1998,Hypki2013}. The MOCCA code simulates and follows the long-term dynamical evolution of spherically symmetric stellar clusters, based on H\'enon's Monte Carlo method \citep[and references therein for details about MOCCA code]{Henon1971,Stodolkiewicz1982,Stodolkiewicz1986,Giersz2013}, together with stellar and binary evolution and strong interactions. In the original version of the code, prescriptions from the SSE/BSE codes \citep{Hurley2000,Hurley2002} are used to follow stellar and binary evolutions, whereas the FEWBODY code \citep{Fregeau2004} handles the strong interactions (binary-binary and binary-single). Finally, escaping stars from tidally limited clusters are treated as described in \cite{Fukushige2000}. 

However, in the version presented in this paper,  named MOCCA-C, only the relaxation process has been included \citep{Henon1971}, and the part of the MOCCA code related to the relaxation process was translated from Fortran to C language. This version is specially designed to be easily integrated into the astrophysical multipurpose software environment (AMUSE\footnote{https://amusecode.github.io/}; \citealt{Pelupessy2013,PortegiesZwart2009,PortegiesZwart2013,PortegiesZwart2018}). 

AMUSE provides a large set of simulation codes and a uniform interface for different kinds of simulations. Indeed, the philosophy within AMUSE is to divide a multi-physics problem into single physical domains, with a specific module that is responsible for the evolution of the system state within its physical domain. The AMUSE environment can communicate between the specific codes through interfaces. This means that the codes used  are interchangeable. In this study, the stellar evolution is handled by the SSE \citep{Hurley2000} version present in the AMUSE environment, whereas the relaxation is handled by the MOCCA-C code. For simplicity and for purpose of testing the new code, the dynamical interactions among stars (binary formations, 3- and 4-body interactions interactions and collisions) has not been implemented in this work.

The system's initial conditions, i.e. positions and velocities for stars, have to be produced outside of MOCCA-C code. This can be done using the AMUSE initial condition procedure or an external code, such as McLuster \citep{Kupper2011}. In the AMUSE framework, stars are represented by AMUSE particles, and they can be handled (added, removed, and evolved) thanks to the AMUSE interfaces.

\subsection{MOCCA-C} \label{subsec:MOCCAC}
The MOCCA-C code contains the relaxation component of the original MOCCA code. The system is firstly divided in zones and superzones \citep{Stodolkiewicz1982,Stodolkiewicz1986}, to better represent the relaxation process in different part of the system (central zones are more frequently relaxed than outer zones - each superzone has its own time step, which increases by a factor of two for each successive superzone); in turn, the relaxation process and new position procedure are applied to all stars in the system \citep{Henon1971}. The computation of a complete time-step is divided into different cycles according to the number of superzones. For each of those cycles, the new positions for each star in the superzones are computed. The determination of changes of the system structure due to changes of the mass distribution  \citep{Stodolkiewicz1982} is applied when the position for all objects in the superzone has been calculated. Finally, stars are removed from the system according to the escape criteria. The current version of the code does not include binaries, so only single stars are considered.

The effect of relaxation in the time interval is mimicked by consecutive encounters between two neighbour stars, with an exchange of energy and angular momentum, as described in \citep{Henon1971,Stodolkiewicz1982,Stodolkiewicz1986}. The new stars' positions are selected randomly between $r_{min}$ and $r_{max}$, with $r_{min}$ being the star's orbit pericentre and $r_{max}$ being the smallest value between the  star’s  orbit apocentre and the outermost radius of the superzone, with probability inversely proportional to the radial velocity $v_r$ at each orbit position. For the outermost superzone in the system, $r_{max}$ is the smallest value between the apocentre distance and the limiting radius ($r_{limit}$), that is set to twice the escape radius ($r_{escape}$). The procedure used to randomly determine the new position is described in \cite{Henon1971}.

Similarly to the standard definitions in NBODY7 code and other version of these codes \citep{Aarseth2012}, the escape radius is set to twice the tidal radius ($r_{tidal}$) for tidally limited clusters, and to ten times the actual $R_h$ for isolated clusters. Alternatively, the escape criteria can be selected from among the following:
\begin{itemize}
    \item distant escape criterion: stars are removed only if their positions are greater than $r_{escape}$;
    \item tidally limited clusters: the removal of bound stars with energy greater than $E_{crit}$ (tidal binding energy) is not instantaneous, but time delayed.  The probability of escape is computed according to the prescription given in  \cite{Fukushige2000}.
\end{itemize}

In MOCCA-C, the stars' orbit is determined from the potential at the beginning of each time step, meanwhile the velocities are estimated at the end, that is after the relaxation step and the new position determination. This inconsistency will lead to a small energy flow. Indeed, the kinetic energy of the stars is altered by the time dependence of the potential, caused by sudden changes  of the stars’ positions inside the system. The kinetic energy corrections are calculated and applied according to the prescription given in \cite{Stodolkiewicz1982}. This procedure will be referred to hereinafter as kinetic energy adjustments due to potential changes in time.

\section{Initial gas conditions and gas expulsion} \label{sec:gasEvo}
The complex physical processes involved in the hydrodynamics of gas-removal from embedded cluster makes it difficult to obtain detailed time evolution of the gas dispersal. In past works, a simplistic analytic representation for the gas expulsion had been used, and the same representation has been used in this work. 

The gas is treated as an external potential to the system \citep[and reference therein]{Lada1984, Kroupa2001-2, Banerjee2013,Banerjee2014,Banerjee_2018b}. The cluster (stars and gas) has been modeled with a Plummer distribution \citep{Plummer1911} for stars, embedded in a spherically symmetric external potential generated by the initial gas. Even though the spatial distribution of gas particles can be different from that of stars, in this work we used the same distributions for both gas and stars. The study for different spatial distributions between the gas and stars as in \cite{Shukirgaliyev2017,Shukirgaliyev2021} will be conducted in the future.

The gas expulsion has been modelled with an exponential decaying function \citep{Banerjee2013}, 
\begin{equation} \label{equation:gas-decay}
M_g \,(t) = \begin{cases} M_g\, (0), & \mbox{if }t \leq \tau_{delay} \\ M_g\, (0) \,\,\, \exp \left( -\frac{t-\tau_{delay}}{\tau_g}\right)   & \mbox{if } t > \tau_{delay}, \end{cases}    
\end{equation}
where $M_g\,(0)$ is the initial mass of the gas, $\tau_g$ is the timescale for gas removal, and $\tau_{delay}$ is the delay time for gas removal. The timescale of gas expulsion is simply given by $\tau_g = R_h\,(0) / v_g$, with $R_h\,(0)$ as the initial half-mass radius of the system, and $v_g$ being the sound speed with which gas expands and becomes removed. The value of $v_g \approx 10$ km/s, the typical sound-speed in a ionized hydrogen region, has been used in this paper (more details can be found in \citealt{Banerjee2013,Banerjee_2018b}). The initial total gas mass has been given by 
\begin{equation*}
M_g\,(0) = M_{cl}\,(0)\,\, \left( \frac{1}{\epsilon} - 1 \right),
\end{equation*}

with $M_{cl}\,(0)$ as the initial total mass of the stars, and $\epsilon$ as the SFE. The total number of gas particles was set to be the same as the number of stars. For our test models, a value of $\epsilon = 0.333$ has been used \citep{Banerjee_2018b}. The gas evolution and expulsion has been treated in the AMUSE environment. During the gas expulsion phase, the gas particles are treated as point mass particles, with mass evolution described by Eq. \ref{equation:gas-decay}. During this evolution, the gas particles' positions are not changed. In future works, the spatial evolution of the gas particles will be included. Finally, the gas and stars' potential have been determined separately during the gas expulsion phase. The potential associated with the gas particles has been interpolated at the stars’ positions taking into account the time evolution of the mass of the gas particles. This contribution has been added to the stars' potentials, determined from the total stars' mass interior to the stars' positions, that is
\begin{equation*}
    u_{star,i} = -G \left( \frac{M_{star,i}}{r_{star,i}} + \sum_{k=i+1}^N \frac{m_{star,k}}{r_{star,k}} \right),
\end{equation*}

where G is the gravitational constant, $u_{star,i}$ and $r_{star,i}$ are the i-th star's potential and position, and $M_{star,i}$ is the total stars' mass interior to $r_{star,i}$.

\subsection{Star cluster evolution phases}
The time evolution of an embedded cluster can be divided into three phases: gas expulsion, violent relaxation and evolution governed by the relaxation process. In the following we will describe the evolution schemes used for each phase. 

\subsubsection{Gas expulsion phase}
The gas expulsion happens on very short time scales. For example, for $R_h = 1$ pc, $\tau_g = 0.1$ Myr (assuming $v_g = 10$ km/s). In order to get reasonable resolution for this phase, the time step was set to be about 10 times smaller than $\tau_g$. At the beginning of each time step, the total mass and potential energy of gas has been updated according to Eq. \ref{equation:gas-decay}. Due to the very small time step, the system has been divided into only one superzone. Successively, the new positions and then the new potential for each star is calculated. Stars whose energy $E > 0$, are treated as unbound, and the procedure used to find their position will be described in \ref{subsec:unbound-stars}. Instead, since for bound stars the time step is much shorter than the crossing time, a proper sampling of the orbit is not possible according to the physical principles behind the Monte Carlo method. For this reason, during this phase the relaxation process has been switched off (kinetic energy between stars is not exchanged). Similarly, no kinetic energy adjustments due to potential changes in time has been applied (as explained at the end of Sec. \ref{subsec:MOCCAC}). The changes of the potential due to gas removal are dominant.

The new positions for bound stars is picked randomly by sampling the orbit. However, the procedure to calculate the bound star's movement along their orbits cannot properly respond to very fast potential changes due to gas expulsion. For stars with energy slightly smaller than zero, the apocenter distance can be very large, implying that the new position can be picked further than the distance the star can travel in the time step. Not taking this into account would lead to the too fast escaping of stars and dissolving of the system. To solve this problem, the maximum distance $r_{max}$ a star can reach is increased at each time step according to the distance the star can travel during the time step $dt$, that is 
\begin{equation} \label{equation:rmax}
r_{max; i,n+1} = r_{max; i,n} + v_{r,n} \cdot dt,
\end{equation}
 where $r_{max; i,n}$ is the maximum distance allowed for the i-th star at time step $n$, $v_{r,n}$ is the radial velocity of the star, and $r_{max; i,0} = v_{r,0} \cdot dt$. This procedure can lead to an artificial delay in the system expansion, since the new position will be always smaller than $r_{max}$. Again, the probability to pick a random position in the bound orbit in a time-step is inversely proportional to the radial velocity $v_r$.
 
 The gas expulsion phase time scale is determined by $\tau_g$, and it is, in general, shorter than 1 Myr. Moreover, a pre-gas expulsion phase can be added by setting a value for $\tau_{delay}$ different from zero. During this phase, the model is evolved according to the procedure described above, with a time step of $\tau_{delay} / 2.0$. This time-step has been chosen only from technical reasons to have a minimum time resolution of this phase. For the chosen value of $\tau_g$ and the initial cluster parameters used in this work, the pre-gas and gas expulsion phases are in general short compared to the mass segregation time scale. For this reason, we assumed that the effects of two-body relaxation are negligible over this time scale. This assumption should not strongly influence the model evolution.

\subsubsection{Violent relaxation phase}
Just after the end of the gas expulsion phase, the system experiences a violent relaxation phase, which brings the system to equilibrium. Indeed, after an early phase during which the half-mass radius oscillates, $R_h$  will eventually settle to an equilibrium value equal to roughly four times the initial half-mass radius  \citep[for $SFE = 0.333$;][]{Lada1984}. The time needed for the system to adjust is directly proportional to the half-mass radius crossing time ($t_{cross,Rh}$ $ = 2\cdot R_h / \sigma_{Rh}$, with $\sigma_{Rh}$ being the velocity dispersion at $R_h$) at the system's maximum extension, which is satisfied at the end of the gas expulsion phase. According to \cite{Lynden-Bell1967}, the time duration of the violent relaxation phase is on the order of a few orbital periods. A value of 4 times the $t_{cross,Rh}$ at the end of the gas expulsion phase was assumed in this work.

The time step during the violent relaxation phase was on the order of $0.2\,\,Myr$. As for the gas expulsion phase, the time step is too small for a proper orbit sample. For this reason, the same treatment described above has been applied, with the system being divided into only one superzone. Both, the relaxation process and the kinetic energy adjustments due to potential changes in time have been switched off during this phase too. Also, the determination of the maximum distance a star can reach as described in Eq. \ref{equation:rmax} has been imposed when determining the new star position. Because of the relatively long time span of this phase, the lack of relaxation and mass segregation in this phase can have an impact on the system expansion, and the spatial structure of the innermost region of the cluster. This can be important for more compact GCs. In future work we plan to introduce relaxation processes during this phase and explore their effects which can be particularly important  for very massive stars.

\subsubsection{Relaxation}
When the violent relaxation phase has ended, the standard Monte Carlo procedure is applied, as described in Sec.  \ref{subsec:MOCCAC}. The time step used in this phase can be $\geq 1.0 \,\,Myr$. To properly account for the cluster mass distribution, the system is divided into at least three superzones.

\subsection{Unbound stars} \label{subsec:unbound-stars}
In the MOCCA code stars with binding energy $E>0$ are removed immediately. This escape criteria is correct only when the time scale to travel across the system is smaller than the overall model time step (usually around $2 - 5$ Myr) so that stars actually have time to travel the system and be expelled during this time step.

In contrast, during the gas expulsion and violent relaxation phase, the time step is too small, and the escaping stars do not have enough time to travel outside the system within the time step. This will lead to some inconsistencies, because the system would too quickly remove escaping stars that would not actually reach the escape radius distance. So the total number of stars bound to the system would decrease faster. Accordingly, the tidal radius would also decrease resulting in smaller Lagrangian radii and finally faster cluster dissolution.

For this reason, stars with $E>0$ are not immediately removed. Instead,  new  positions  and  velocities for such stars are calculated by an approximated orbit integration.  In this treatment, it is assumed that an unbound object can move outward and only radially, i.e. $r_{i,n+1} = r_{i,n} + |v_r| \cdot dt$, with $v_r = v$ and $v$ being the total velocity of the object. From energy conservation $E_{i,n+1} = E_{i,n}$, we obtain the radial velocity at the end of the time step, $v_{r,n+1} =  \sqrt{\,2.0 \cdot E_{i,n} - 2.0 \cdot u_{i,n+1}}$, with $u_{i,n+1}$ giving the potential of the star at position $r_{i,n+1}$.

This procedure has been applied for the entire simulation (through all different phases). It is important to underline that unbound stars are also considered in the changes in energy and angular momentum estimations  during the relaxation process, but they are not taken into account during the kinetic energy adjustments due to potential changes in time. The unbound stars do actually move for small distances in the system during one time step, and the potential changes are not important compared to changes introduced by the relaxation process.

\subsection{Pros and cons of the gas expulsion treatment}
The gas removal phase and the following violent relaxation phase have been introduced for the first time in the MOCCA code. The procedure of gas expulsion  described in this paper was based on a few assumptions that may lead to differences when compared to N-body results for the two initial phases. The movement of bound and unbound stars in the system has been adjusted in order to handle the small cluster evolution time-steps used during those phases. Additionally, the procedure used for the unbound stars movement is simplistic, lacking proper orbit integration. The lack of primordial and dynamically formed binaries, together with old fashion stellar evolution, can lead to slower mass loss and system expansion. The lack of relaxation and mass segregation during the initial phases (particularly during the violent relaxation phase) as explained before, can lead to a slightly slower system expansion and core-collapse. Summing up all those simplifications can lead to some differences in the Lagrangian radii, particularly for the outermost and innermost ones.

We focused our attention on the spatial structure and mass of the cluster after the violent relaxation and aimed at obtaiing results matching as closely as possible those from N-body simulations. In this way we can assume that the final and longest phase of the cluster evolution we simulate will follow the evolution of star cluster in the same way as N-body simulations.

\section{Simulations and Results} \label{sec:SimAndRes}
To test our code, a comparison with N-body simulations is needed. For this purpose, we run two different sets of simulations: the first one, a training test, consists of 4 simulations that have been run with NBODY7 \citep{Aarseth2012}; alternatively, in the second set we try to reproduce the results showed in \cite{Banerjee2013}.

The initial conditions have been generated using a newly updated version\footnote{The updated version of the code can be found in Github: \url{https://github.com/agostinolev/mcluster}.} of the McLuster code\footnote{The original version of McLuster can be found at \url{https://github.com/ahwkuepper/mcluster}.}\citep{Kupper2011}. For more details see Appendix \ref{Appendix}.

\subsection{Comparison with new N-body simulations} \label{sec:NewNbody}
The training test consists of four simulations with two different numbers of particles $N=[100000$, $200000]$ and two initial half-mass radii $R_h = [0.5$, $1.0]$ pc. The positions and velocities of stars for each model were selected according to the Plummer model \citep{Plummer1911}. The \cite{Kroupa2001} initial mass function (IMF)  raging from $0.08 \,\,M_\odot$ to $100 \,\, M_\odot$ was applied. For all those models, no primordial binaries were included. The models were run up to $50$ Myr. The escape criteria radius $r_{escape}$ was set to a constant value of $100$ pc (with $r_{limit}$ set to $50$ pc). The gas expulsion time-scales have been set to $\tau_g = 0.05$ and $0.1$ Myr for models with $R_h = 0.5$ and $1.0$ pc, respectively.The time delay for gas expulsion was set to $0.1$ Myr for all models.

\begin{figure}
    \centering
    \begin{subfigure}{0.5\textwidth}
       \centering
        \includegraphics[width=\linewidth]{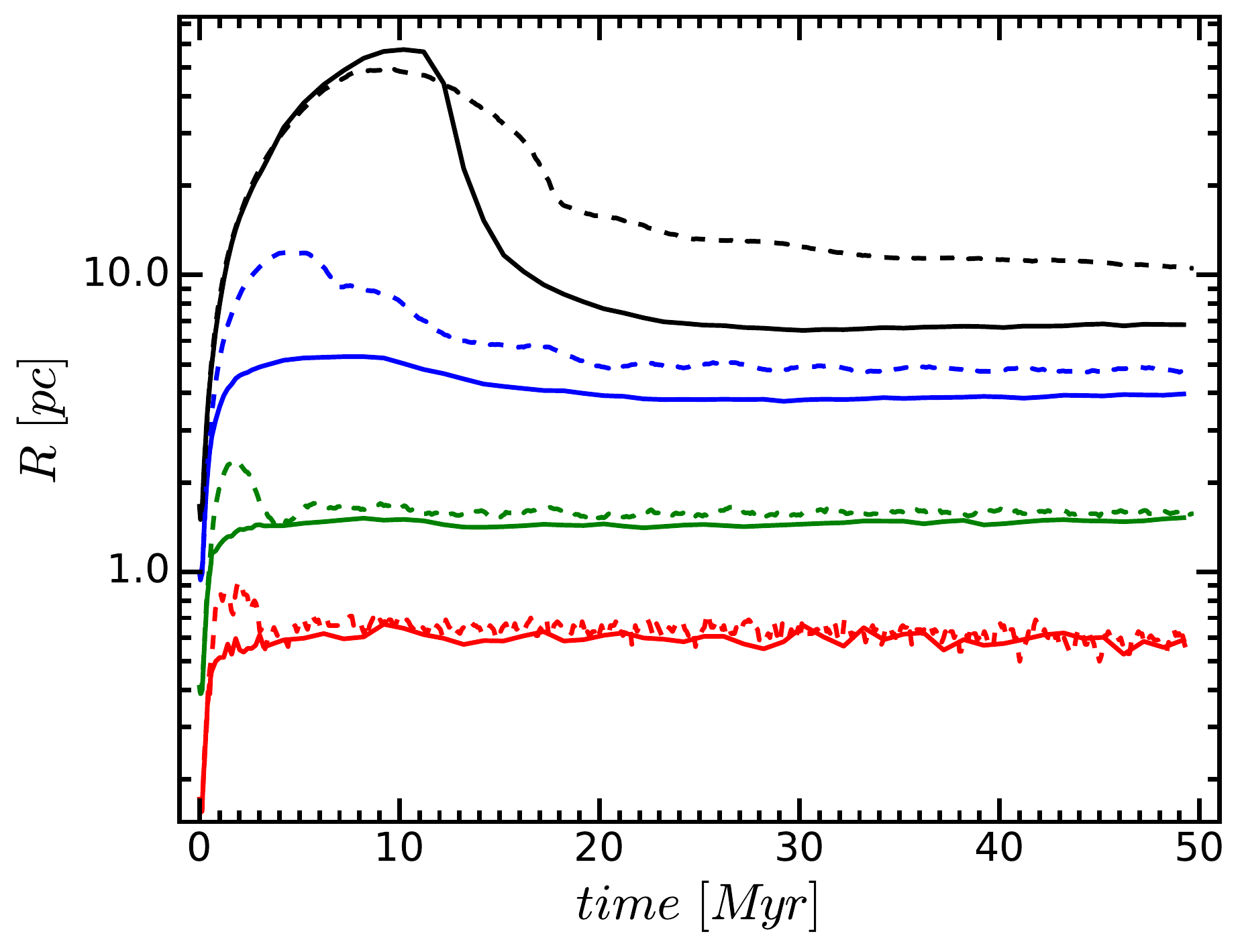}
    \end{subfigure}
    \begin{subfigure}{0.5\textwidth}
       \centering
        \includegraphics[width=\linewidth]{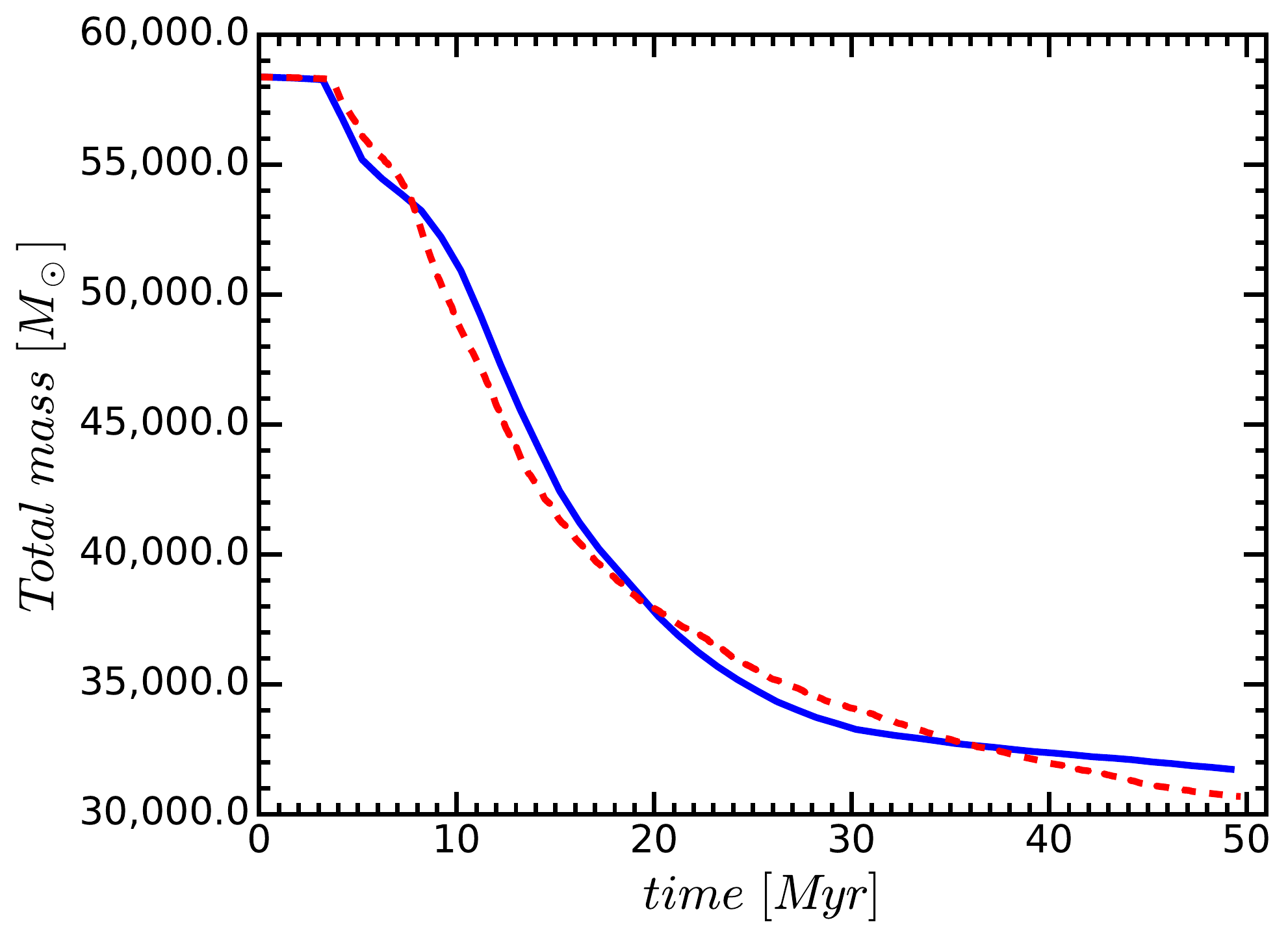}
    \end{subfigure}
    \begin{subfigure}{0.5\textwidth}
       \centering
        \includegraphics[width=\linewidth]{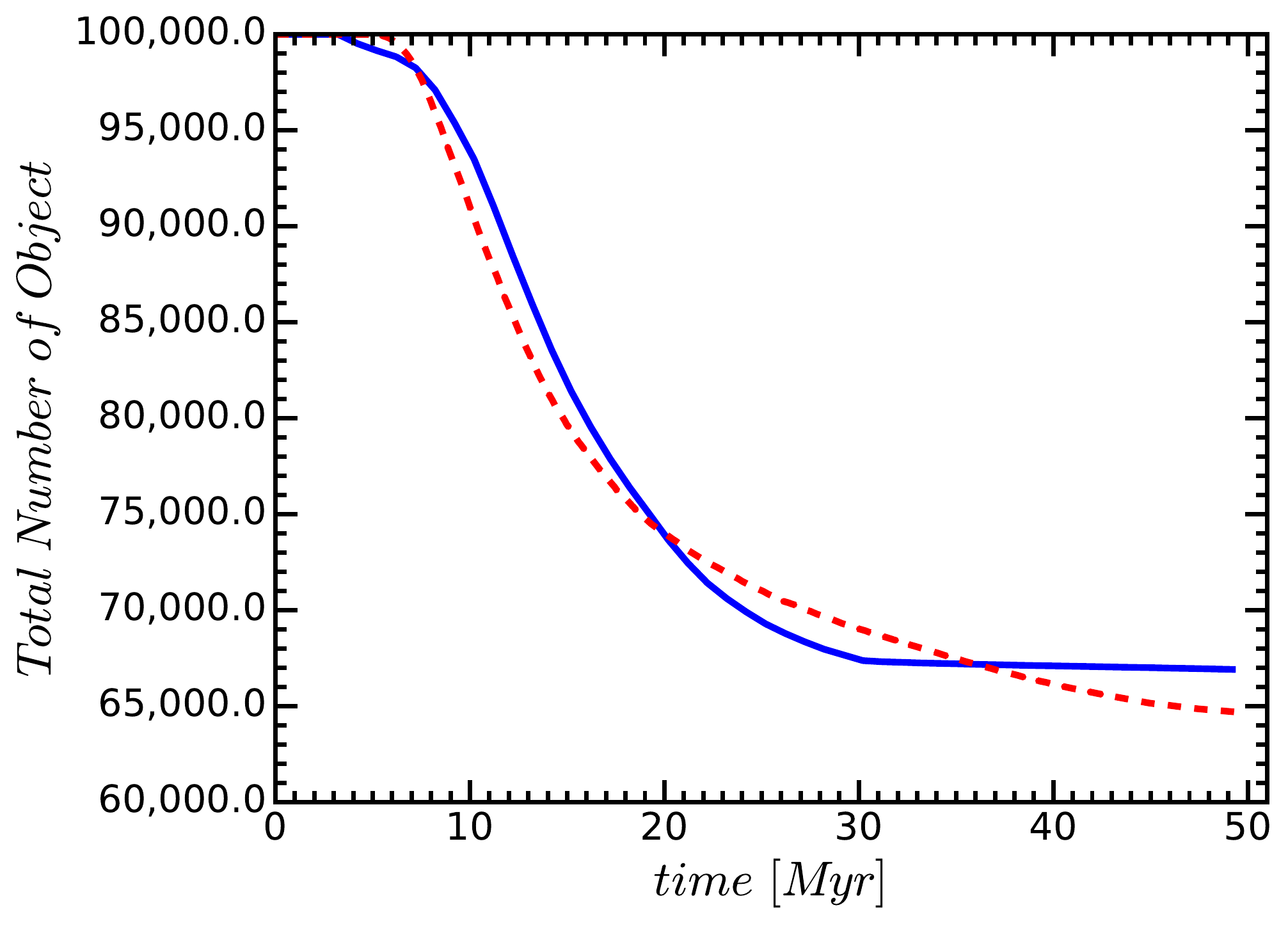}
    \end{subfigure}
    \caption{Lagrangian radii (top), total mass (middle) and total number of bound objects (bottom) evolution for the simulation with $N=100000$ and $R_h=1.0\,\,pc$. The continuous  and dashed lines correspond to the MOCCA-C and N-body results, respectively. In the upper figure, the curves, from bottom to top, correspond to 1, 10, 50 and 75$\%$ Lagrangian radii evolution.The gas expulsion phase for this model ended at 0.58 Myr, while the violent relaxation phase ended at $\sim 6.0$ Myr.}
    \label{Fig:S-N100k-rh1.0}
\end{figure}

\begin{figure}
    \centering
    \begin{subfigure}{0.5\textwidth}
       \centering
        \includegraphics[width=\linewidth]{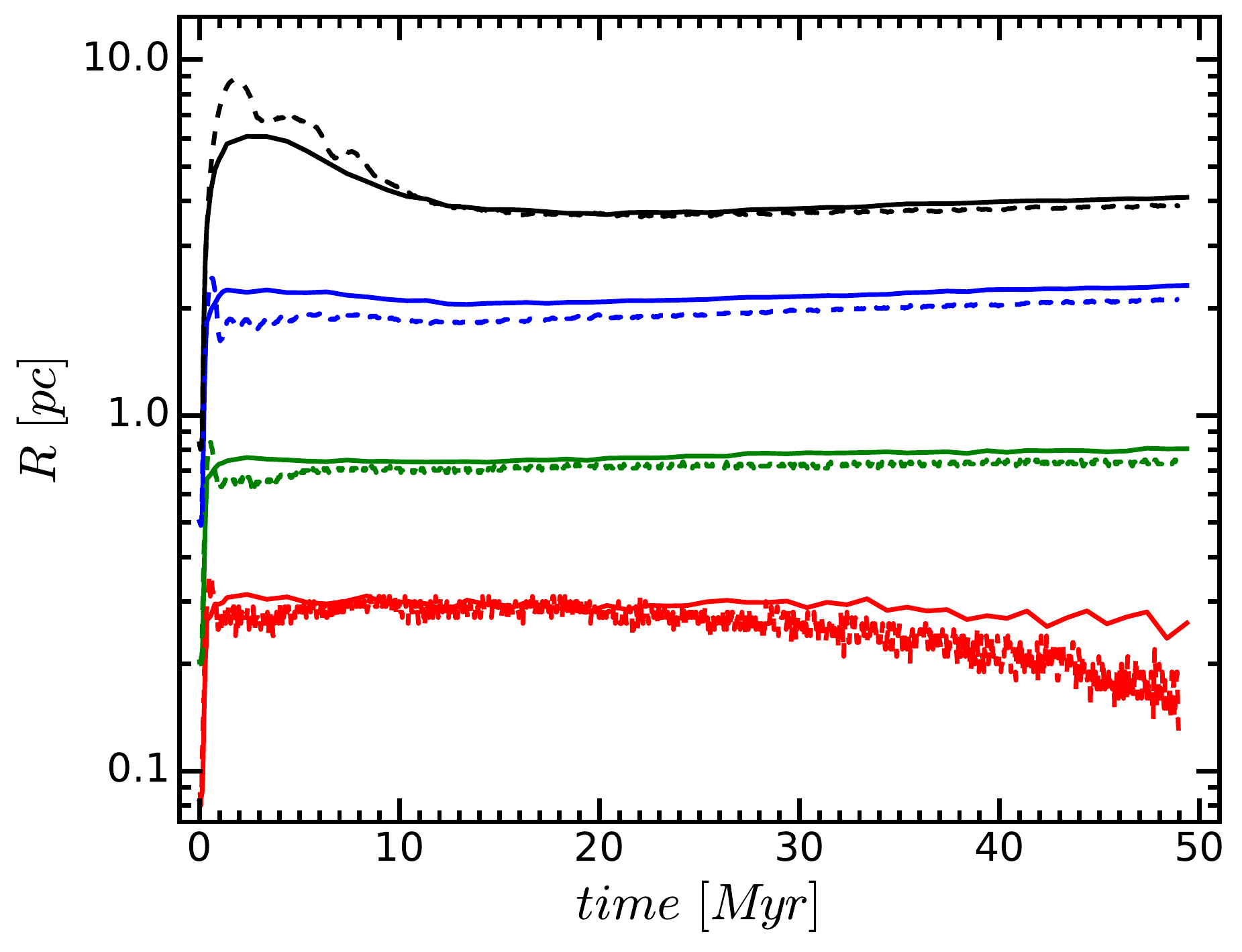}
    \end{subfigure}
    \begin{subfigure}{0.5\textwidth}
       \centering
        \includegraphics[width=\linewidth]{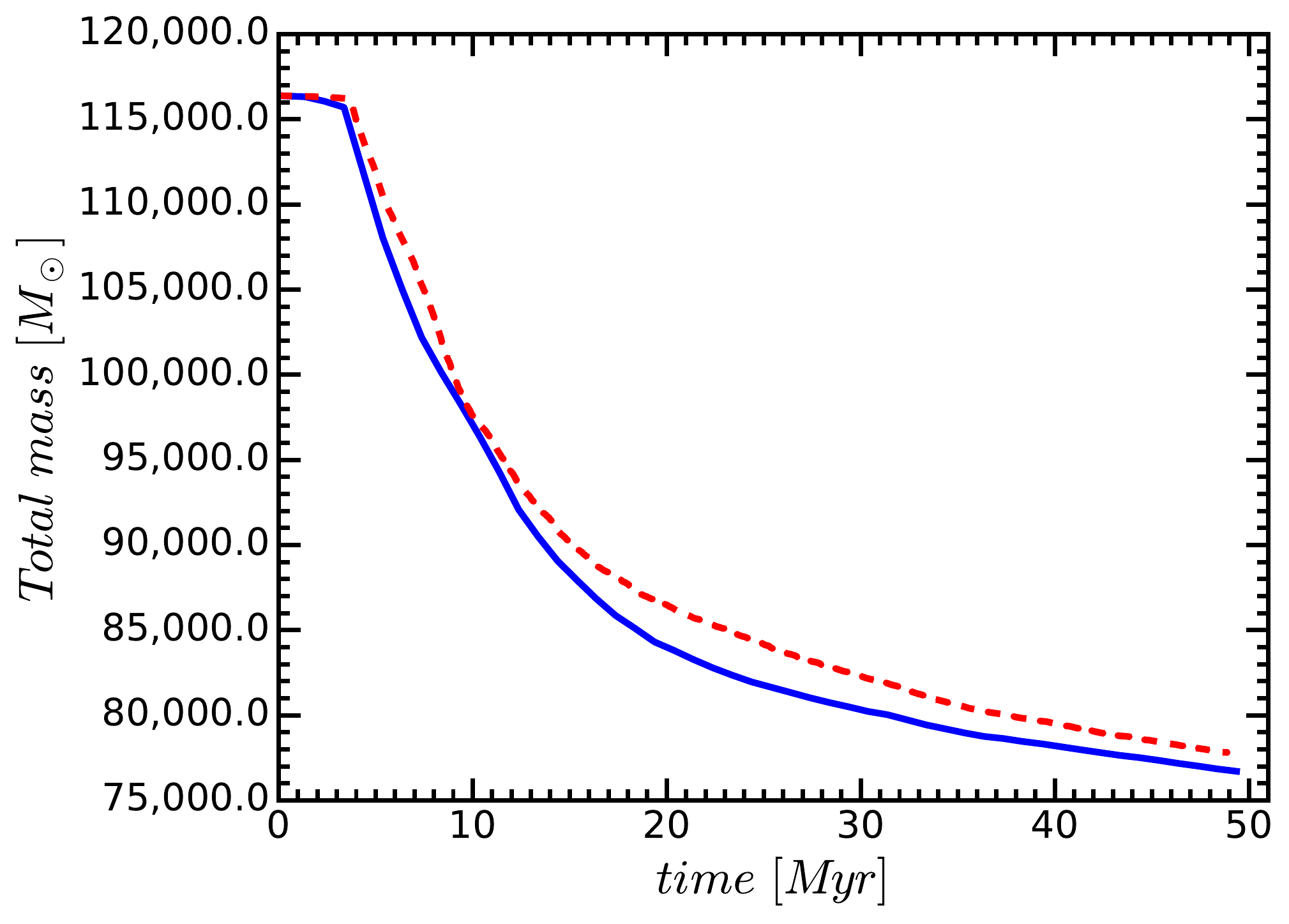}
    \end{subfigure}
    \begin{subfigure}{0.5\textwidth}
       \centering
        \includegraphics[width=\linewidth]{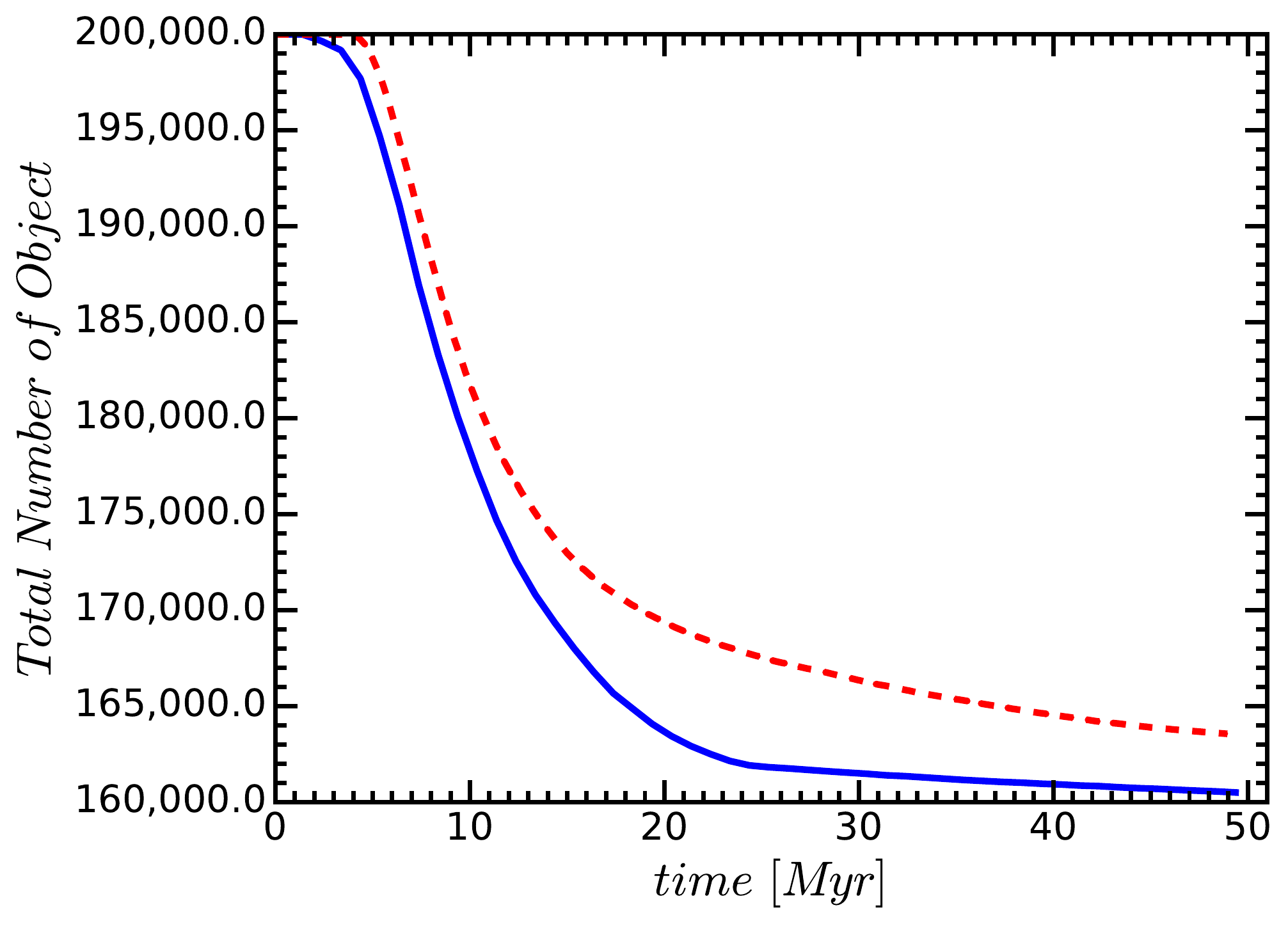}
    \end{subfigure}
    \caption{Lagrangian radii (top), total mass (middle) and total number of bound objects (bottom) evolution for the simulation with $N=200000$ and $R_h=0.5\,\,pc$. The continuous  and dashed lines correspond to the MOCCA-C and N-body results, respectively. In the upper figure, the curves, from bottom to top, correspond to 1, 10, 50 and 75$\%$ Lagrangian radii evolution. The gas expulsion phase for this model ended at 0.34 Myr, while the violent relaxation phase ended at $\sim 5.0$ Myr.}
    \label{Fig:S-N200k-rh0.5}
\end{figure}

The Lagrangian radii, the evolution of the total mass and of the total number of objects for two of the models are reported in Fig. \ref{Fig:S-N100k-rh1.0}  and \ref{Fig:S-N200k-rh0.5}. The other two models show similar evolution. The gas expulsion phase for the model reported in Fig. \ref{Fig:S-N100k-rh1.0} ended at 0.58 Myr, and the violent relaxation phase at $\sim 6$ Myr. Instead, for the model reported in Fig. \ref{Fig:S-N200k-rh0.5}, the gas expulsion phase ended at 0.34 Myr, and the violent relaxation phase at $\sim 5$ Myr. One can clearly see that our prescription can reasonably well reproduce the N-body results, with some differences in the outermost Lagrangian radii, and innermost Lagrangian radii for the model with initial $R_h = 0.5\,\,pc$.
To quantify the differences between the MOCCA-C and N-Body results for the Lagrangian radii, the total mass, and the total number of bound objects, we integrate the areas below the individual curves for each quantity and calculate the ratio of the difference between these areas to the area under the line for the N-body simulations. The differences shown in Fig. \ref{Fig:S-N100k-rh1.0} are of the order of 15\%, 17\%, 20\% and 11\% for 1, 10, 50 and 75\% Lagrangian radii respectively at the end of gas expulsion phase, and of the order of 4\%, 4\%, 18\% and 56\% at the end of the simulation. The differences in mass and in total number of bound object are both of the order of 1\% at the end of gas expulsion phase, and of 3\% at the end of the simulation. A better comparison is shown in Fig. \ref{Fig:S-N200k-rh0.5}. The differences for the  1, 10, 50 and 75\% Lagrangian radii  at the end of gas expulsion phase are of the order of 11\%, 9\%, 4\% and 9\% respectively, with values of 25\%, 9\%, 9\%, and 5\% respectively at the end of the simulation. The differences in mass and in total number of bound object are both of the order of 1\% at the end of gas expulsion phase, and of 1-2\% at the end of the simulation. The evolution of the total mass and total number of bound objects is reproduced reasonably well, with an important part of the mass loss being connected to the number of star escaping the system.

Those differences can be explained by differences present in the cluster structure connected to the approximate treatment of the star movement and relaxation process. The differences seen in the outermost region of Fig. \ref{Fig:S-N100k-rh1.0} are related to the approximate treatment for the unbound stars that lead to an important mass loss. Instead, the differences in Fig. \ref{Fig:S-N200k-rh0.5} are important in the central region of the system. In this case, the lack of the formation of binaries is responsible for causing our model to have a delayed core collapsed when compared to the N-body simulation. Also, during the violent relaxation phase, the energy exchange among stars can be important and induce mass segregation in the N-body simulation that is instead absent in the MOCCA simulation since the effects of relaxation are not included in this phase.  Furthermore, the differences in the stellar evolution prescription may play some role in those dissimilarities: in the current version of MOCCA-C we use the older version of the stellar evolution prescription from \cite{Hurley2000} and \cite{Hurley2000}, while the NBODY prescription already includes  the most up-to-date, similar to the Level C in \cite{Kamlah2021}. Mass loss due to stellar evolution for the same models with the old and new stellar evolution prescriptions are about 14\% and 17\%, respectively. Likewise, the average BH masses are 20.9$ M_\odot$ and 25.6$ M_\odot$, for the old and new stellar evolution prescription respectively. Consequently, larger mass loss will lead to a larger cluster expansion (i.e.,  larger Lagrangian radii, particularly for the outermost ones) in the models with the new stellar evolution. On the other hand, the larger BH masses found in the models with the new stellar evolution will lead to a more rapid evolution towards core collapse and denser systems. Due to the strong initial cluster expansion, the binary formation efficiency is relatively low, with only 1 or 2 binaries formed during the simulated time span in the N-body simulations.
 
\subsection{Comparison with previous works}
As second comparison, we tried to reproduce the results shown in \cite{Banerjee2013}. In that paper, only the initial few $Myr$ for the GC R136 and NGC 3603 were simulated. The initial total masses for those models were $M_{cl} \,(0) = 10^5\,\,M_\odot$ and $M_{cl} \,(0) = 1.3 \times 10^3\,\,M_\odot$, respectively. The initial half-mass radius followed the \cite{Marks2012} relationship, giving $R_h = 0.45$ pc and $0.34$ pc, respectively. For these simulations we adopt the same escape criterion used in \cite{Banerjee2013} and set the escape radius equal to 10 times the current half-mass radius value. The \cite{Kroupa2001} IMF was used, with the most massive stars set according to the \cite{Weidner2004} relationship.

In our simulations, we used the \cite{Kroupa2001} IMF, with a minimum mass of $0.08 \,\,M_\odot$ and a maximum mass of $150$ and $50 \,\, M_\odot$, for GC R136 and NGC 3603 respectively. In order to reproduce the initial total mass for those clusters, a initial total number of stars of $170000$ and $22000$ were used, respectively. The statistical fluctuations for the low-N MOCCA-C models can introduce substantial noise in the global system parameters determination. To reduce the noise, each simulation has been repeated 20 times with a different random seed. All relevant quantities for the model evolution have been computed from the mean of the 20 models. The time delay for gas expulsion was set to $0.0$ and $0.6$ Myr for NGC 3603 and R136, respectively. 

\begin{figure}
    \centering
    \includegraphics[width=\linewidth]{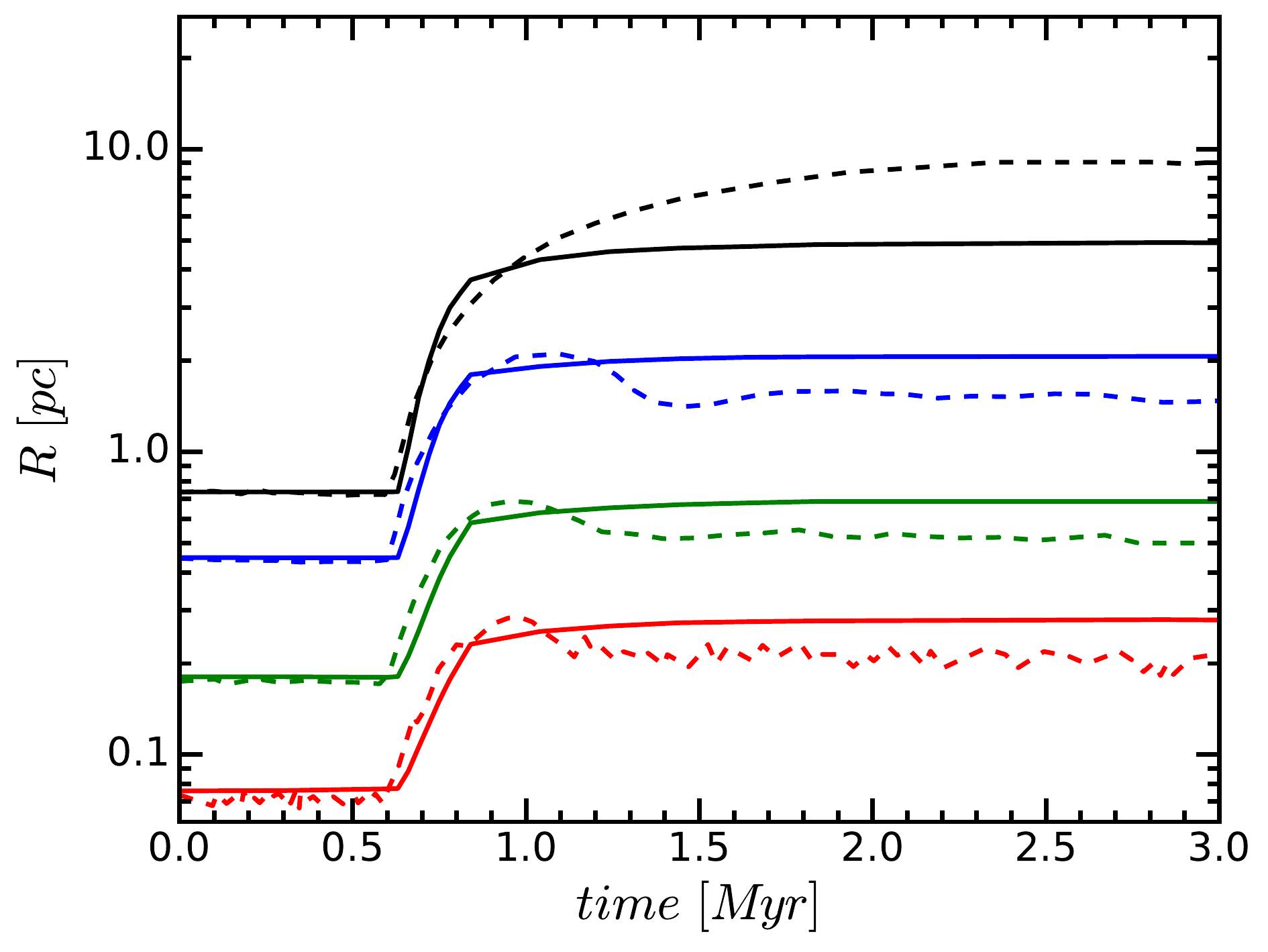}
    \caption{Lagrangian radii evolution for N-body in dashed lines and MOCCA-C in continuous lines for the model with $M_{cl} \,(0) = 1.0 \cdot 10^5\,\,M_\odot$, and $N(0) = 170000$ (the model for R136).  The curves, from bottom to top, correspond to 1, 10, 50 and 75$\%$ Lagrangian radii evolution. The N-body model is from  Fig. 1 of \protect\cite{Banerjee2013}.}
    \label{Fig:R136}
\end{figure}

\begin{figure}
    \centering
    \includegraphics[width=\linewidth]{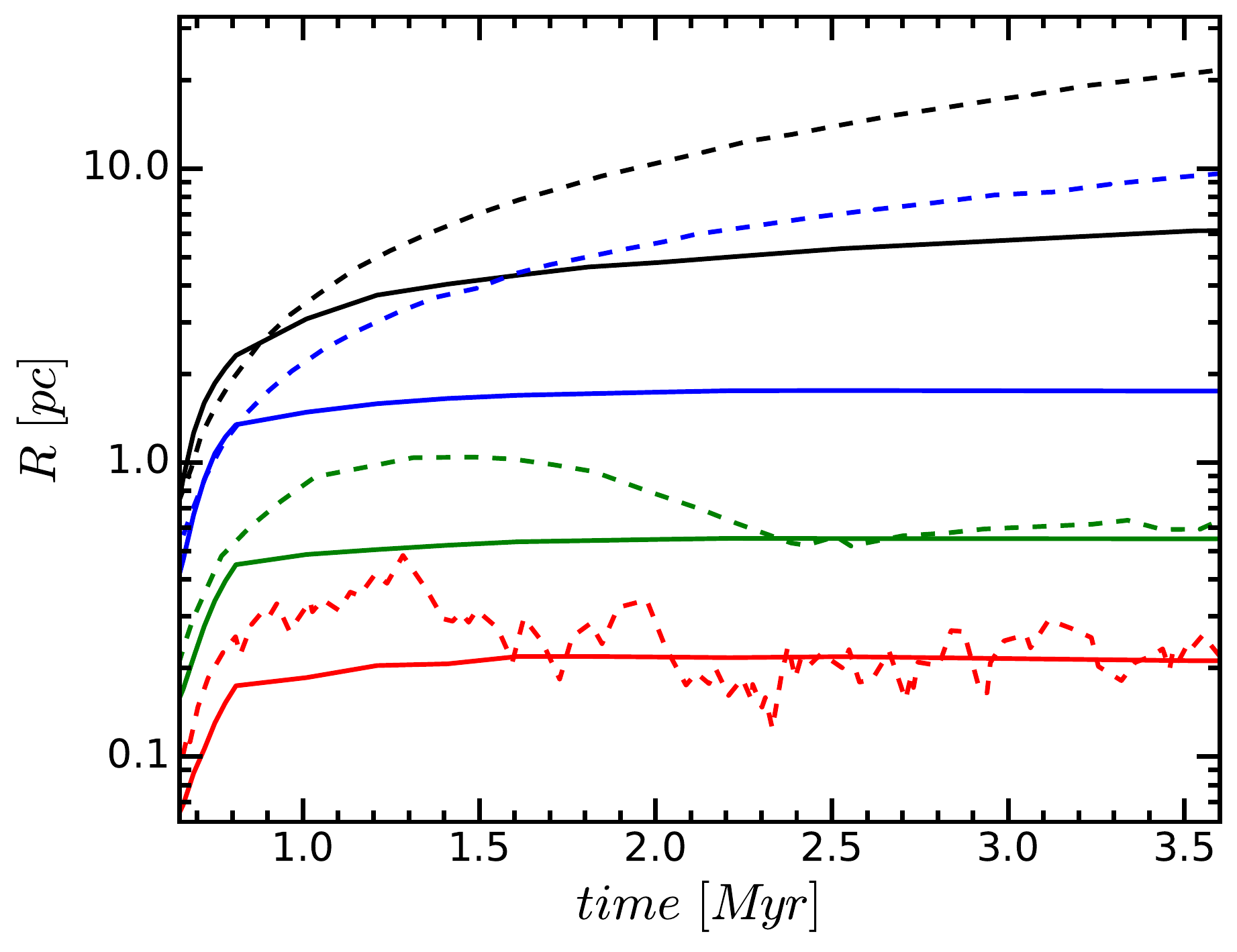}
    \caption{Lagrangian radii evolution for N-body in dashed lines and MOCCA-C in continuous lines for the model with $M_{cl} \,(0) = 1.3 \cdot 10^3\,\,M_\odot$, and $N(0) = 22000$ (the model for NGC 3603). The curves, from bottom to top, correspond to 1, 10, 50 and 75$\%$ Lagrangian radii evolution. The N-body model is from  Fig. 4 of \protect\cite{Banerjee2013}.}
        \label{Fig:NGC3603}
\end{figure}

The time evolution of the Lagrangian radii for R136 and NGC 3603 are reported in Fig. \ref{Fig:R136} and \ref{Fig:NGC3603}, respectively. Despite the small time-steps and the short evolution, the system expansion is relatively well accounted for, at least for the R136 model. For NGC 3603 the too low number of objects, being at the limit of the Monte Carlo method,  introduces important statistical fluctuations. Indeed, for small N the treatment of unbound objects is probably too simplified and leads to observed differences for larger Lagrangian radii. As for the training set, small differences in the innermost radii are visible. The duration of the gas expulsion and violent relaxation can be relatively long, up to several Myr, and during that time the massive stars can exchange energy and angular momentum. This will lead to some mass segregation (as seen in N-body simulations), despite the fact that violent relaxation does not depend on stellar mass. On the other hand, the lack of mass segregation in our treatment can lead to a smaller system expansion. Finally, the different starting point showed in Fig. \ref{Fig:NGC3603} for smaller Lagrangian radii is because the initial time starts at 0.6 Myr (the time of gas expulsion), following the original figure from \cite{Banerjee2013}. Using the same procedure described above for Fig. \ref{Fig:S-N100k-rh1.0}  and \ref{Fig:S-N200k-rh0.5}, we determined the differences between the MOCCA-C and N-Body results. Our results for R136 differ by 20\%, 23\%, 27\% and 77\% for 1, 10, 50 and 75\% Lagrangian radii respectively; instead, the results for NGC3603 differ by $\sim 5\%$ for 1 and 10\% Lagrangian radii, and are 3 and 2 times smaller compared to the N-Body results for 50 and 75\% Lagrangian radii respectively. These values are relatively large, since these simulations span a limited time range including only the gas expulsion phase and the beginning of the violent relaxation phase; as shown in Fig.  \ref{Fig:S-N100k-rh1.0} and Fig. \ref{Fig:S-N200k-rh0.5}, these are the phases when the  differences between the MOCCA and the NBODY simulations are larger.

Despite some of the physical simplifying assumptions adopted in the Monte Carlo procedure, our results show that there is a general satisfactory agreement between the MOCCA-C and N-body simulations and that the Mone Carlo simulations are able to capture the main aspects of these early evolutionary phases.
%The agreement between the MOCCA-C and N-body simulations showed so far (with new and previous models) is much better than we expected, taking into account the weak points of the physical assumptions behind the Monte Carlo procedure.

\subsection{MOCCA-C models of large-N clusters}
The main advantage of MOCCA over NBODY is in its ability to model systems with large N much more rapidly while still producing reliable results \citep{Giersz2013, Wang2016,Kamlah2021}. Indeed, at most several hours were needed to run the models presented in Sec. \ref{sec:NewNbody}, in contrast to the roughly two weeks needed for NBODY7 to run the same models. MOCCA simulations thus allow to extend the study the evolution of clusters with large N during the gas expulsion phase. We present here the results of a set of simulations for models following the same prescription for gas removal adopted in the previous section. For the test simulations, which show the correct operation of the code and its effectiveness, we chose Plummer models with $N=5\cdot 10^5$ objects, $R_h = 0.5, \,\,1.0,\,\,3.0 \,\,,\,\,6.0$ pc. The $R_h$ have been chosen so that the first two models are tidally underfilling, the third one is tidally filling at the end of the gas expulsion, and the last model to be nearly initially  tidally filling. The gas expulsion time-scales $\tau_g$  have been set to  $0.05$, $0.1$, $0.3$ and $0.6$ Myr, and the time delay for gas expulsion was set to $0.1$ Myr. The tidal radius was set to $60.0$ pc. The IMF was from \cite{Kroupa2001}, ranging from $0.08\,\,M_\odot$ to $150\,\,M_\odot$. For each model, we set the SFE $\epsilon = 1.0,\,\,0.7,\,\,0.5,\,\,0.4,\,\,0.333,\,\,0.2$ and $0.1$. Each model has been run up to $2$ Gyr. In this preliminary work, as said before, no binaries are allowed (neither initially nor dynamically). For this reason, the evolution of the models is stopped when core collapsed has been reached.

With those simulations, we aimed to investigate the importance of the SFE on the evolution and survival of the system. Indeed, as showed in previous works (both theoretical \citealt{Kroupa2001-2,Baumgardt2007,Geyer2001}, and observational \citealt{Lada2003,Elmegreen2000}), a minimal SFE of 0.333 is needed to form a bound and gas-free cluster in dynamical equilibrium. So far, this has been tested for small N models only. 

\begin{figure*}
    \centering
    \includegraphics[width=\textwidth]{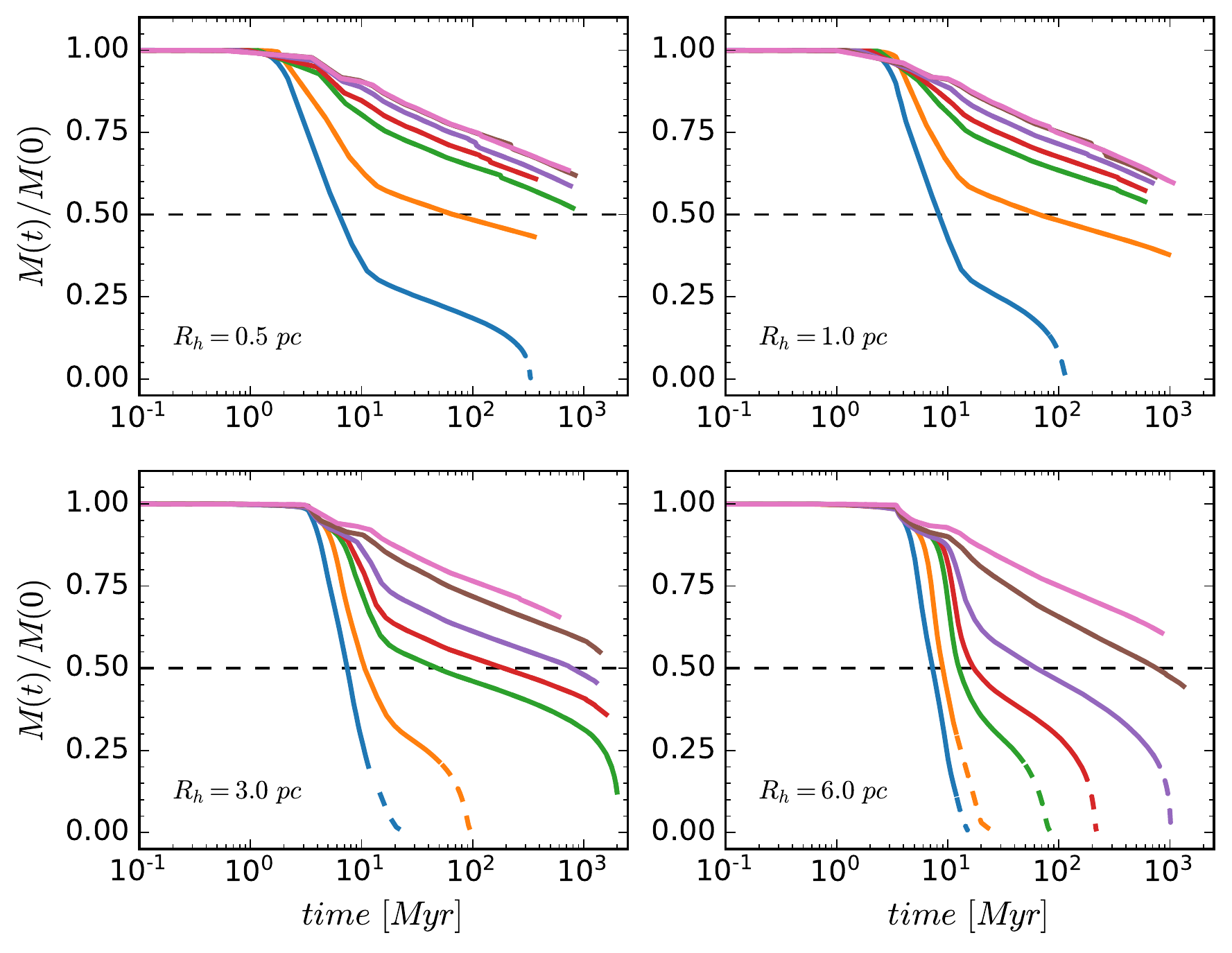}
     \caption{MOCCA-C evolution of the ratio between the actual mass and the initial mass for the model run as a function of SFE. The upper left panel shows runs with $R_h = 0.5\,\,pc$, the upper right with  $R_h = 1.0\,\,pc$, the lower left with $R_h = 3.0\,\,pc$ and the lower right with $R_h = 6.0\,\,pc$. In each panel, the lines from left to right show the models with SFE = $0.1$, $0.2$, $0.333$, $0.4$, $0.5$, $0.6$, $0.7$, $1.0$. The dashed part of the lines shows the times for which the tidal disruption of the model has already begun. The plots start at time $0.1\,\,Myr$, which is at the beginning of gas expulsion.}
     \label{Fig:RunModels_mass}
\end{figure*}

\begin{figure*}
    \centering
    \includegraphics[width=\textwidth]{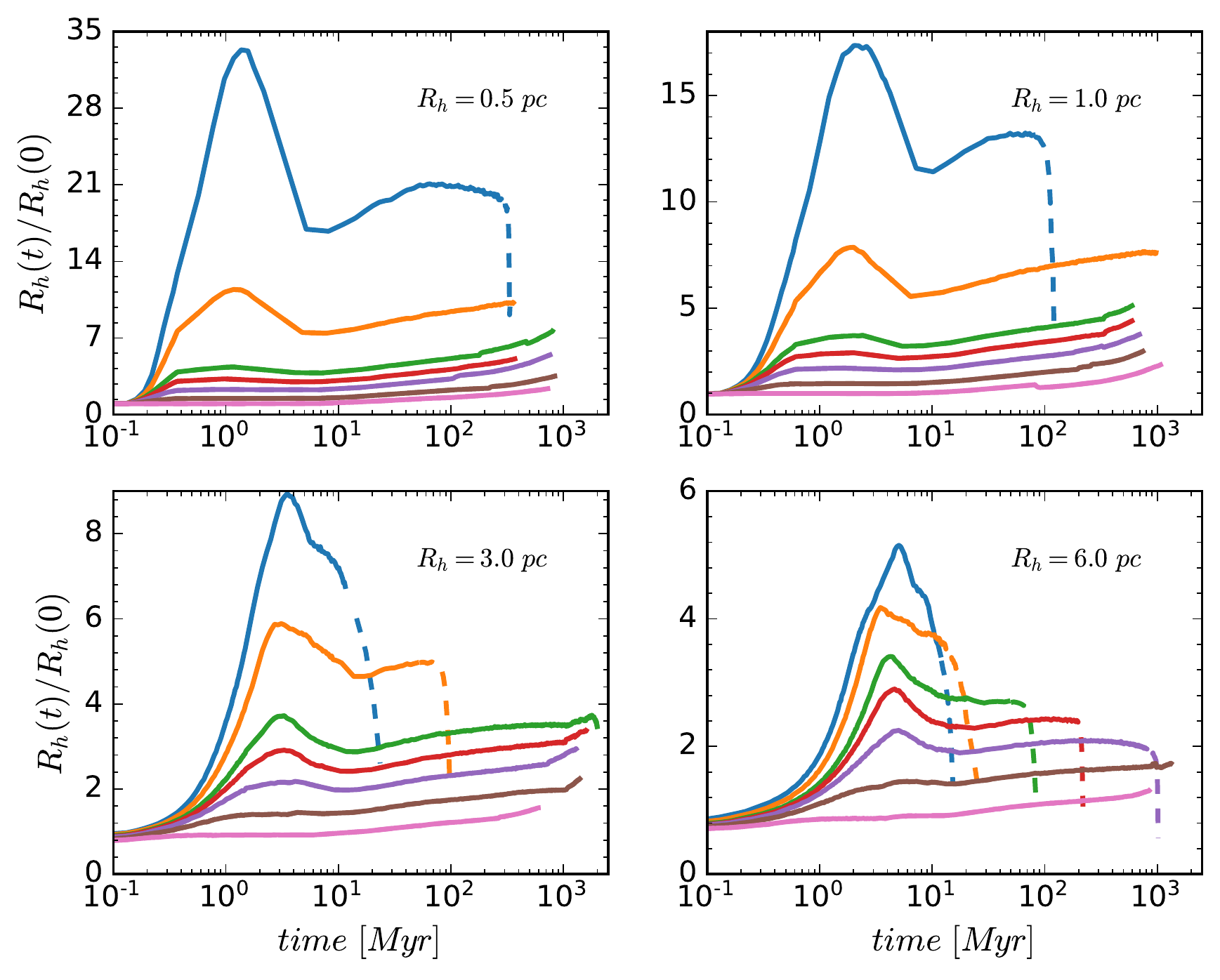}
     \caption{Same as in Fig. \ref{Fig:RunModels_mass}, but showing the evolution of the ratio between the actual and the initial half mass radii.  In each panel, the lines from top to bottom show the models with SFE = $0.1$, $0.2$, $0.333$, $0.4$, $0.5$, $0.6$, $0.7$, $1.0$.}
     \label{Fig:RunModels_rh}
\end{figure*}

The evolution of total mass (scaled by the initial values) is showed in Fig. \ref{Fig:RunModels_mass}. The importance of the initial state of the system (tidally filling or underfilling) is visible in Table \ref{Table:RunModel}, where the ratio between the final mass and the initial mass for each model is reported. The models that were initially tidally filling are unable to survive the embedded gas phase, being dissolved  for $\epsilon \leq 0.5$. Indeed, only for the cases with $\epsilon = 0.7$ and $1.0$ the models survive, even though an important percentage of the initial mass had been removed, with the final half-mass radius being $\sim 10.0$ pc. Models that were initially tidally undefilling ($R_h = 0.5$, $1.0$ and $3.0$ pc) reached an advance state of evolution, with roughly half (or more) of the initial mass. Interestingly, no model with $\epsilon = 0.1$ has survived. However, models that were strongly tidally underfilling can survive the gas expulsion with $\epsilon \geq 0.2$. Star clusters that were formed as tidally filling or only slightly tidally underfilling can survive the gas expulsion only if $\epsilon \geq 0.7$ and can be observed as star clusters advanced in age. This means that star clusters with the observationally suggested values of $\epsilon = 0.333$ have to be formed as tidally underfilled. On the other hand, clusters that are very strongly tidally underfilled can survive under a very low SFE, equal to about $0.2$. Also, in Figs. \ref{Fig:RunModels_mass} and \ref{Fig:RunModels_rh}, we reported in dashed lines the time when the dissolution of the system started. As discussed in other works \citep{Fukushige1995,Contenta2015,Giersz2019}, the dissolution of the system happens on a dynamical time-scale, when the system loses its dynamical equilibrium and  will not undergo core collapse.

The evolution of the ratio between the actual and the initial  half-mass radii for those models are showed in Fig. \ref{Fig:RunModels_rh}. The value of  $R_h$  at the end of the gas expulsion strongly depends on the SFE and the initial $R_h$. Indeed, for initial values of $R_h = 0.5,\,\,1.0,\,\,3.0$ and $6.0$ pc, the $R_h$ at the end of gas expulsion is $\sim 34$, $\sim 17$, $\sim 10$ and $\sim 5$ times larger than the initial value for $\epsilon = 0.1$, respectively. The drastic expansion of $R_h$ is clearly visible for small SFE in all models, and for $\epsilon \leq 0.7$ for the $R_h=6.0$ pc model. In contrast, for $\epsilon = 0.333$ this value is $\sim 3-4$ times larger, independent of the initial $R_h$.  Taking into account the value of the tidal radius, a Plummer model (that can be roughly modelled with a King profile with $w_0 = 5.0 - 6.0$) is tidally filling when its $R_h \simeq 8.0$ pc. Therefore, the model with an initial value of $R_h = 3.0$ pc and $\epsilon = 0.333$ will become tidally filling at the end of the gas expulsion. Finally, at the end of the violent relaxation, the half mass radius shows a drastic drop, with the actual half mass radius at that time being only a few pc. The further evolution of $R_h$ is principally governed by the relaxation process and mass loss.


\begin{table}
\centering
\begin{tabular}{ccccc}
$\epsilon$ & $R_{h}=0.5$ & $R_{h}=1.0$ & $R_{h}=3.0$ & $R_{h}=6.0$ \\
\hline
0.1 & - (335.2) & - (121.2) & - (23.5) & - (12.3) \\
0.2 & 0.4 (358.8) & 0.4 (978.4) & - (94.7) & - (26.1) \\
0.333 & 0.5 (799.2) & 0.5 (599.4) & 0.1 (1997.7) & - (74.1) \\
0.4 & 0.6 (370.0) & 0.6 (599.0) & 0.4 (1597.5) & - (218.9) \\
0.5 & 0.6 (759.8) & 0.6 (697.6) & 0.5 (1299.1) & - (1019.5) \\
0.7 & 0.6 (837.6) & 0.6 (748.2) & 0.5 (1399.3) & 0.4 (1327.5) \\
1.0 & 0.6 (729.5) & 0.6 (1069.0) & 0.7 (597.1) & 0.6 (849.8) \\
 \hline 
\end{tabular}

\caption{The ratio between the final and initial masses for model runs with different initial $R_h$ and $\epsilon$. Models that were disrupted are reported with a dash, meanwhile all other models undergo a core collapse at the end of the simulations. In brackets the time (in $Myr$) for which the models stop is reported. The gas expulsion time-scales $\tau_g$  have been set to  $0.05$, $0.1$, $0.3$ and $0.6$ Myr,  for the models with $R_h = 0.5$, $1.0$, $3.0$ and $6.0$ pc, respectively.}
\label{Table:RunModel}
\end{table}

\section{Discussions and Conclusions} \label{sec:Conclusions}
We have introduced and studied for the first time the embedded gas phase and gas removal phases in the evolution of star clusters with the Monte Carlo MOCCA code. By using the MOCCA code, it is possible to extend the study of these evolutionary phases to systems with a number of stars much larger than those allowed by direct N-body simulations.

For the study presented in this paper, we have developed a simplified version of MOCCA running within the AMUSE environment. The investigation presented in this paper is a pilot study to show that it is possible to follow the evolution of star clusters during the gas expulsion phase in the Monte Carlo framework and that we can produce results in general good agreement with those of N-Body simulations.

The new version of MOCCA introduced here includes a new treatment for unbound stars necessary to reproduce the early cluster expansion, while additional dynamical processes will need to be added in future developments of the code. 
In particular, to reproduce more realistic models of globular cluster evolution and survival after the gas expulsion phase, a better treatment of relaxation and mass segregation in the violent relaxation phase is necessary. Moreover, while a constant SFE and a simplistic treatment of the gas expulsion have been assumed for this initial study, future work will include a more realistic treatment of SFE and gas expulsion.

Despite the differences between the recipes adopted for stellar evolution in the N-body and the Monte Carlo simulations, we do not find significant differences in the dynamics of the systems studied. Instead, the largest differences are seen from the comparison with the models from \cite{Banerjee2013} and can be explained by the small number of stars which are at the limit of applicability of the Monte Carlo method.

New models have been run in order to investigate the importance  of the  SFE and tidal field  on  the  evolution  and  survival  of  large $N$  system. In the assumed SFE and the gas removal prescription, models with SFE $= 0.1$ dissolve within $10-100$  Myr. Models that were initially tidally filling ($R_h = 6.0$ pc) were able to survive only for large SFE ($\geq 0.7$). Instead, all other models, formed a bound system and survived the gas expulsion phase.

In \cite{Baumgardt2007}, the authors performed a large set of N-body simulations with different SFEs, strengths of the tidal field and gas expulsion time-scales. Each simulation consisted of models with $N=20000$ equal-mass stars, and no stellar evolution was included. As a result, models that were tidally underfilling and with $\epsilon = 0.1$ were able to create bound clusters after the gas expulsion, but only if the tidal fields were weak ($R_h/R_{tidal} = 0.01-0.03$) and the gas was removed slowly ($\tau_g/t_{cross} = 10$). Also, the authors find that no bound clusters were formed for tidally filling models. Our results are in agreement with the ones shown in \cite{Baumgardt2007}, considering that the models we run in this paper have a faster gas removal ($\tau_g/t_{cross} \lesssim 5$). Nevertheless, our study present more realistic models including the effects of a sprectrum of stellar masses and those of stellar evolution. The presence of massive stars in the system and the stellar evolution in the first Myr  can be important for the long-term cluster survival. Indeed, the lack of any massive objects, can lead to a different mass loss efficiency and dynamical interaction which in turn leads to not deep enough core collapses and slower dissolutions of the cluster. Their findings still approximately hold also for larger and multi-mass clusters.

The clusters studied in \cite{Brinkmann2017} also survived the gas expulsion phase even in a strong tidal field. The authors run several NBODY7 models with initial $R_h$ equal to $0.1$ pc, $0.3$ pc and $0.5$ pc, $M_{cl}$  ranging from $5 \cdot 10^3 $  to $5 \cdot 10^4 M_\odot$, and multi-mass IMF according to \cite{Kroupa2013}. Similarly as in this work, a Plummer model for the gas and stellar component, and an identical treatment for the gas depletion was used. The models presented in that work are strongly underfilling, and their clusters survived the gas expulsion, as we also see in this paper. The authors also found that the final bound fraction can be sensitive to the relation $\tau_g/t_{cross}$, implying a different final bound mass fraction for clusters with the same initial central density. Additionally, similar to our results, their findings  show that for fixed gas expulsion velocity, the SFE plays an important role for the survival and bound mass of the system, with larger SFE leading to larger bound fractions. Finally, we find that the stellar evolution can influence the bound mass fraction, due to the different re-virialization timescales after the gas expulsion, with clusters without stellar evolution having higher bound masses than the models with stellar evolution. These results are additionally supported by the MOCCA-C models. This gives us confidence that the gas removal and violent relaxation treatments proposed in this paper can be safely, but with some care, used to simulate large-N embedded systems.

Future work will focus on introducing the embedded gas phase and the procedures introduced in this paper to the standalone MOCCA code. Indeed, the presence and formation of binaries in the system, such as the dynamical interactions and collisions among stars and binaries strongly influence the evolution of the system. In very dense systems (with initial $R_h \sim 0.1$ pc), the collision and dynamical interactions can be important from the very beginning of system evolution, including the violent relaxation phase. Further studies will need to be performed, which will take binaries and interactions into account. Interactions and collisions among stars and binaries, particularly for massive stars and binaries, may strongly influence the early cluster evolution. In fact, the presence of very massive stars, massive BHs and binaries (primordial and dynamically formed) in the system, together with the energy they may release in the system through dynamical interaction may have a significant impact on the dynamics and survival of the cluster both during its early and long-term evolutionary phases. These aspects will be studied in a series of future papers.

\section*{Acknowledgements}

MG and AL were partially supported by the Polish National Science Center (NCN) through the grant UMO-2016/23/B/ST9/02732. SB acknowledges support from the Deutsche Forschungsgemeinschaft (DFG; German Research Foundation) through the individual research grant “The dynamics of stellar-mass black holes in dense stellar systems and their role in gravitational-wave generation” (BA 4281/6-1; PI: S. Banerjee). 

\subsubsection*{Energy consumption of these simulations}
We run MOCCA-C for about 2000 single-core CPU hours. This results in about 37.25 kWh of electricity (\url{http://green-algorithms.org/}) being consumed by the CAMK supercomputer. With our estimate of the proportion of green electricity used, this process produces $\sim$ 30 kg CO2,
which is comparable to driving a car about 170 km.
%%%%%%%%%%%%%%%%%%%%%%%%%%%%%%%%%%%%%%%%%%%%%%%%%%
\section*{Data Availability}

The data underlying this article will be shared on reasonable request to the corresponding author. The AMUSE code is available for download via GitHub at \url{https://amusecode.org}. The codes used in this paper are: Python \citep{vanRossum}, matplotlib \citep{Hunter}, numpy \citep{Oliphant}, MPI \citep{Gropp2,Gropp1}, SSE/BSE \citep{Hurley2000,Hurley2002}.


\bibliographystyle{mnras}
\bibliography{biblio}

%%%%%%%%%%%%%%%%%%%%%%%%%%%%%%%%%%%%%%%%%%%%%%%%%%

%%%%%%%%%%%%%%%%% APPENDICES %%%%%%%%%%%%%%%%%%%%%

\appendix

\section{McLuster, a new implementation} \label[Appendix A]{Appendix}
The original version of star cluster initial model generator McLuster, developed by \cite{Kupper2011}, has been updated, to better link the generated initial conditions with the MOCCA code.

The main update of the new implementation has been an upgrade to the existing procedure for generating multiple stellar populations (with a maximum of 10) as initial conditions. For a single stellar population, the positions and velocities of stars for each population can be generated according to different models: Plummer \citep{Plummer1911}, King \citep{King1966}, Subr \citep{Subr2008}. Instead, for multiple stellar populations the velocities of stars are obtained by solving the Jeans equation for dynamical equilibrium. Indeed, for two or more populations which are individually in virial equilibrium, their combination may not be. Moreover, the final system does not follow the distribution of the single population models (i.e., two Plummer models do not sum to a Plummer model). In order to establish virial equilibrium, the velocities of the stars have been modified accordingly. The mass density and the mass profiles for the entire system are determined and used to determine the velocity dispersion profile for the model solving the Jeans equation. The velocity of each star is then obtain from a normal distribution, with the standard deviation equal to the local velocity dispersion. Additionally, stellar and binary evolution are allowed for each population.

Also, it is now possible to apply a semi-major axis uniform distribution in $\log(a)$ for low mass stars and a distribution for the orbital period for high mass stars based on \cite{Sana2012, Oh2015}. The modified pre-main sequence eigenevolution (\cite{Belloni2017}; the original procedure described in \citet{Kroupa1995b,Kroupa2013}) has been added as an option in generating binary properties. Generally, the pre-main sequence eigenevolution procedure would modify the mass of stars that compose the binaries. For this reason, the total mass of the system and the conversion parameters (from physical units to Nbody) may differ. In order to avoid such an error, the order in which the procedures are called has been modified, with the primordial binary property determination called before the determination of the conversion parameters (opposite to the original version, where the primordial binary property procedure is called last).

A configuration file has been included into the code, in order to modify the initial conditions parameter in a simpler way. The format of the output of the initial model generated can be used for N-body and/or MOCCA simulations.

Finally, the possibility to evaluate the potential energy in spherical symmetry has been added. As an $O(N)$ algorithm this drastically speeds up the code for models containing millions of objects. This add-on is essential for the MOCCA initial conditions, since the primary assumption in Monte Carlo codes is spherical symmetry of the system.

In table \ref{Table:mcluster}, the initial parameters in the configuration file are reported. From left to right, we reported the name of the parameter, the description of the parameter and some additional notes.

\bsp	% typesetting comment
\label{lastpage}
\clearpage
\begin{table}
    \centering
    \begin{tabular}{c|c|l}
    Parameter & Description & Extra note \\
    \hline
    n & Initial number of objects  & $n = n_{singles} + n_{binaries}$ \\
    \hline 
    $fb$ & Primoridal binary fraction & $n_{binaries} = fb \cdot n$ \\
    \hline 
    \multirow{4}{*}{initialModel}  & \multirow{4}{*}{Initial density distribution} & 0 - Homogeneous sphere \\ 
    & & 1 - Plummer \citep{Plummer1911} \\ 
    & & 2 - King \citep{King1966}\\ 
    & & 3 - Subr \citep{Subr2008}\\
    \hline 
    $w_0$ & King model parameter & Values between $1.0 - 12.0$ \\
    \hline 
    S & Mass segregation parameter &Values between $0.0 - 1.0$ \\
    \hline
    fractal & Fractal dimensions & Values between $0.0 - 3.0$ (3.0 not fractal) \\
    \hline 
    qvir  & Virial ratio& $qvir > 0.5  $: expanding; $qvir = 0.5 $: equilibrium; $qvir < 0.5 $: collapsing\\
    \hline 
    \multirow{4}{*}{mfunc} & \multirow{4}{*}{Stellar mass function} & 0 - Equal masses\\ 
    & &  1 - \cite{Kroupa2001} IMF\\ 
    & &  2 - Multi-power law\\ 
    & & 3 - L3 IMF \citep{Maschberger2013}\\
    \hline 
    \multirow{4}{*}{pairing} & \multirow{4}{*}{Pairing of binary components} & 0 - Random pairing\\
    & & 1 - Ordered pairing for components with masses $M > 5 M_\odot$\\
    & & 2 - Random but separate pairing for components with masses $M > 5 M_\odot$\\
    & & 3 - Uniform distribution of mass ratio for $M > 5 M_\odot$, random pairing for $M \leq 5 M_\odot$ \\
    \hline 
    \multirow{7}{*}{adis} & \multirow{7}{*}{Semi-major axis distribution} & 0 - uniform distribution in log(a)\\
    & & 1 - Lognormal distribution distribution\\
    & & 2 - \cite{Kroupa1995a}  period distribution \\
    & &  3 - \cite{Kroupa1995a}  period distribution for $M < 5 M_\odot$; \cite{Sana2012} for $M > 5 M_\odot$\\
    & & 4 - Flat uniform distribution between amin and amax\\
    & & 5  - \cite{Duquennoy1991} period distribution \\
    & & 6 - Uniform distribution in log(a) for $M < 5 M_\odot$; \cite{Sana2012} for $M > 5 M_\odot$\\
    \hline 
    \multirow{3}{*}{eigen} & \multirow{3}{*}{Eigenevolution}   & 0 - Off\\
    & &  1 - \cite{Kroupa1995b} eigenevolution\\
    & & 2 - \cite{Kroupa2013}, rewieved in \cite{Belloni2017}\\
    \hline 
    amin & Min. binary semi-major axis & Value in $R_\odot$\\
    \hline 
    amax & Max. binary semi-major axis & Value in $R_\odot$\\
    \hline 
    tf & Tidal field & No tidal field or point mass galaxy \\
    \hline 
    rbar & Tidal radius &  Value  in parsec\\
    \hline 
    rh\_mcl  & Half mass radius  & Value  for the whole system, in parsec\\
    \hline 
    \multirow{2}{*}{conc\_pop} & \multirow{2}{*}{Concentration radius parameter} & Defined as $Rh_i/Rh_1$, the ratio between the half-mass \\
    & & radii of the i-th and the first generation \\
    \hline 
    \multirow{2}{*}{potential\_energy} & \multirow{2}{*}{Potential energy evaluation} & Potential energy evaluated in spherical symmetry \\
    & & sum of gravitational potential for every object \\
    \hline 
    epoch & Age of population  & Value in Myr\\
    \hline 
    zini & Initial metallicity & $zini_{\odot} = 0.02$ for Solar metallicity\\
    \hline 
    seedmc & Random number generator & \\
    \hline 
    outputf & Output format & Initial files for MOCCA and/or N-body simulations\\
    \hline 
    check\_en &Make energy check at end &\\
    \hline
    BSE & Activate SSE/BSE   & Swtich on/off stellar/binary evolution\\
    \hline 
    \end{tabular}
    \caption{Mcluster initial parameters.}
    \label{Table:mcluster}
\end{table}
\clearpage


%%%%%%%%%%%%%%%%%%%%%%%%%%%%%%%%%%%%%%%%%%%%%%%%%%

% Don't change these lines

\end{document}